\documentclass[10pt,draftcls,onecolumn]{IEEEtran}

\usepackage{epsf,amssymb,amsfonts,cite,amsbsy}
\usepackage{color}
\usepackage{graphicx}
\usepackage{subfigure}
\usepackage{amsmath}
\usepackage{amssymb}
\usepackage{mathrsfs}
\usepackage{cite}
\usepackage{hyperref}
\usepackage{url}
\usepackage{multirow}
\usepackage{rotating}
\usepackage{setspace}
\usepackage{lipsum}
\usepackage{multicol}
\usepackage{float}
\usepackage{amsthm}

\newtheorem{theorem}{Theorem}

\newtheorem{remark}{Remark}

\DeclareMathOperator*{\Gtrless}{\gtrless}

\DeclareMathOperator{\x}{\mathbf{x}}
\DeclareMathOperator{\X}{\mathbf{X}}
\DeclareMathOperator{\z}{\mathbf{z}}

\DeclareMathOperator{\IR}{\mathbb{R}}

\ifCLASSINFOpdf
\else
\fi

\begin{document}

\title{Fusing Censored Dependent Data for Distributed Detection}

\author{\IEEEauthorblockN{Hao He and Pramod K. Varshney} \\
\IEEEauthorblockA{Department of EECS, Syracuse University, Syracuse, NY 13244, USA\\
Email: hhe02 and varshney@syr.edu}
}
\maketitle
\begin{abstract}
In this paper, we consider a distributed detection problem for a censoring sensor network where each sensor's communication rate is significantly reduced by transmitting only ``informative" observations to the Fusion Center (FC), and censoring those deemed ``uninformative". While the independence of data from censoring sensors is often assumed in previous research, we explore spatial dependence among observations. Our focus is on designing the fusion rule under the Neyman-Pearson (NP) framework that takes into account the spatial dependence among observations. Two transmission scenarios are considered, one where uncensored observations are transmitted directly to the FC and second where they are first quantized and then transmitted to further improve transmission efficiency. Copula-based Generalized Likelihood Ratio Test (GLRT) for censored data is proposed with both continuous and discrete messages received at the FC corresponding to different transmission strategies.  We address the computational issues of the copula-based GLRTs involving multidimensional integrals by presenting more efficient fusion rules, based on the key idea of injecting controlled noise at the FC before fusion. 
Although, the signal-to-noise ratio (SNR) is reduced by introducing controlled noise at the receiver, simulation results demonstrate that the resulting noise-aided fusion approach based on adding artificial noise performs very closely to the exact copula-based GLRTs.
Copula-based GLRTs and their noise-aided counterparts by exploiting the spatial dependence greatly improve detection performance compared with the fusion rule under independence assumption.
\end{abstract}

\IEEEpeerreviewmaketitle
\section{Introduction}
\label{sec:introduction}
Advances in computational capabilities of the constituent sensor nodes inspired a surge of interest in distributed detection, in which the sensors send their locally processed data instead of raw observations to the FC, and the FC makes the final decision according to a certain fusion rule \cite{Tsitsiklis93decentralizeddetection,varshney1997distributed}.  A new transmission-efficient distributed detection framework is considered in \cite{489500,4456697,1413462,1545843}, based on a send/no-send idea. The sensors ``censor" their observations according to a certain mechanism to satisfy the communication rate constraints.  In this process, sensors send their observations to the FC only if they are deemed ``informative". Thus, only a subset of observations are received at the FC for decision making. 
It has been proved that with conditionally independent sensor data, transmission occurs if and only if the local likelihood ratio falls outside of a single ``no-send" interval, under both NP and Bayesian frameworks \cite{489500,4456697}. 

Detection problems with censoring sensors have been investigated from various perspectives, e.g., utilizing sequential detection \cite{Addesso07}, assuming fading channels \cite{Biao05} or under eavesdropper attacks \cite{Willett09}.
In \cite{Tsitsiklis07}, the authors investigated the optimal censoring and transmission strategies by considering an asymptotic criterion involving error exponents in a network where sensors have access to some side information. 
The idea of censoring has also been applied for data reduction for estimation purposes in \cite{CTYu,6093758,yujiao12}. 


Prior research on censoring for distributed detection and estimation has been carried out under the assumption of conditionally independent observations. However, dependence often occurs in practice as the sensors observing the same phenomenon are likely to have spatially dependent observations. The design of the censoring scheme at local sensors and the fusion rule at FC becomes highly complex as a result of dependence among observations.
The effect of dependence on the performance of distributed detection has been investigated recently in \cite{haochen12,30797,Iyengar2012,Hao2012,Hao2013a}. The authors in \cite{Kapnadak11} and \cite{Whipps13} considered physics-based models of spatial correlation and protocol-based communications and constraints, with alternative forms of censoring. 
The detection problem with censoring sensors considered in \cite{Douglas2010,Hao2013}, assumes that spatial dependence among observations is known to the FC. In many practical situations, such information is not available, due to either the intrinsic non-stationarity of the signal \cite{Hao2012} or heterogeneity of the sensing modalities. 

In this work, we consider the fusion of censored data for distributed detection in a heterogeneous sensor network under unknown inter-sensor dependence. To tackle the issue of unknown spatial dependence which can be nonlinear and quite complex, we apply the statistical theory of copulas which has previously been used for hypothesis testing with analog and quantized data in \cite{Iyengar2011,Iyengar2012}. The main contributions of this paper are as follows:
\begin{itemize}
\item A copula-based GLRT for analog censored data is proposed under the NP framework, which, by accommodating unspecified spatial dependence, generalizes our prior study on censoring in \cite{Hao2013}.  For Wireless Sensor Networks (WSNs) with limited transmission resources including power and channel bandwidth, we also consider the scenario where the uncensored measurements are quantized to further reduce the amount of data transmitted from sensors to the FC\cite{1413462,6093758}. 
For the fusion of the discrete data with unknown dependence at the FC a copula-based GLRT is derived. The GLRTs for both analog censored data and quantized-censored data exploit the censoring mechanism and the inter-sensor dependence for improved detection performance. As will be evident later, the two GLRTs are computationally expensive, especially in a large sensor network.
\item To address the computational issue of the copula-based GLRT for analog censored data, an alternative fusion approach is proposed. In this approach, each unreceived message is substituted with the noise generated according to a uniform distribution at the FC. Such approximated fusion rule reduces the computational complexity of the ``exact" copula-based GLRT with very small amount of performance loss.
\item Further, to address the computational issue of the copula-based GLRT for quantized-censored data, a noise-aided GLRT is proposed based on Widrow's theorem of quantization \cite{Widrow-Kollar,Iyengar2012}, in which 
a controlled noise is added to these discrete-valued signals. 
The noise-aided approach greatly simplifies the fusion rule by completely eliminating the necessity of computing multidimensional integrals while achieving comparable detection performance in certain ranges of censoring rate. 
\end{itemize}
The four fusion rules proposed in this paper are able to accommodate different sensor modalities (different marginal PDFs), different individual censoring constraints, and different channel capacities (number of bits that can be transmitted) and, therefore, are applicable to many practical detection problems in heterogeneous sensor networks.

The rest of the paper is organized as follows.
In Section \ref{sec:prob}, we present our model in detail and formulate the problem of censoring. In Section \ref{sec:fusion_copula},  copula theory is introduced and copula-based fusion rules are derived under two different transmission scenarios. In Section \ref{sec:censored}, we propose an alternate fusion rule for analog censored data to address the computational issue.
In Section \ref{sec:quantized-censored}, we propose a computationally efficient fusion rule for quantized-censored data based on Widrow's quantization theorem. Simulation results are provided in Section \ref{sec:sim}. Section \ref{sec:c} includes some concluding remarks and suggestions for future work.

\section{Problem formulation}
\label{sec:prob}
We consider the detection problem in a sensor network where the two hypotheses are denoted by $H_0$ (null) and $H_1$ (target). A total of $N$ sensors are deployed to observe the phenomenon of interest and the Fusion Center (FC) also takes its own observations. We use the random variables $X_{n}$ and $X_{0}$ to respectively denote the observation of sensor $n$ and the observation of the FC at each time instant. It is assumed that the real-valued observations of each sensor and the FC are independent and identically distributed (i.i.d) over time with known probability density function (PDF) $f_n(x_{n}|H_0)$ and $f_n(x_{n}|H_1)$, respectively under hypotheses $H_0$ and $H_1$, i.e., for each sensor $n$ during any time interval $1 \leq l \leq L$
\begin{eqnarray}
f(x_{n1}, \dots, x_{nL}|H_i) = \prod_{l=1}^L f_n(x_{nl}|H_i)
\end{eqnarray}
where $x_{nl}$ denotes the observation at time instant $l$ and $L$ is the length of the decision window. However, spatial dependence exists among sensors' and FC's observations, i.e. for each time instant $l$, 
\begin{eqnarray}
f(x_{0l}, \dots, x_{Nl}|H_i) \neq \prod_{n=0}^N f_n(x_{nl}|H_i)
\end{eqnarray}
and it is not specified.

In a censoring sensor network, each sensor node decides to transmit or not based on a function of its own observation $h_n(x_{nl})$, such as the likelihood ratio function. When $h_n(x_{nl})$ falls in the sending region $R_n^{'}$ the message $u_{nl}$ is transmitted, otherwise nothing is sent. Thus, the censoring operation takes the following form:
 \begin{eqnarray}
\left \{
  \begin{array}{l l}
    h_n(x_{nl}) \in R_n^{'}, & \quad \text{$u_{nl} = {\gamma_n}^{'}(x_{nl})$ is sent} \\
    h_n(x_{nl}) \in \overline{R_n^{'}}, & \quad \text{nothing is sent}
  \end{array} \right.
\label{eq:cen_general}
\end{eqnarray}
where the complement set $\overline{R_n^{'}}$ is the censoring/no-send region and ${\gamma_n}^{'}(\cdot)$ denotes the mapping from observation to message. 

In many detection problems, ``null" ($H_0$), namely the ``normal condition", occurs much more frequently than ``target" ($H_1$), and therefore, the probability of censoring is considered only under $H_0$ in this paper.
In the NP framework, the censoring region $\overline{R_n^{'}}$ of each sensor satisfies an individual censoring rate constraint as follows 
\begin{eqnarray}
P(h_n(X_n) \in \overline{R_n^{'}}|H_0) = \mathit{\beta_n}
\end{eqnarray}
where $0<\mathit{\beta_n}<1$ and 
\begin{eqnarray}
P(h_n(X_n) \in \overline{R_n^{'}}|H_0)=\int \limits_{h_n(x_n) \in \overline{R_n^{'}}} f_n(x_n|H_0) dx_n
\end{eqnarray}
is the probability that sensor $n$ censors its observations under $H_0$. Censoring results in the transmission of the most ``informative" observations under the censoring rate constraint so as to attain the best possible detection performance.  


It has been proved in \cite{489500} that given conditionally independent observations, the optimal $h_n(\cdot)$ and $\gamma_n^{'}(\cdot)$ in the censoring scheme (\ref{eq:cen_general}) are both likelihood ratio functions, i.e., 
\begin{eqnarray}
l_n(x_{nl})=\frac{f_n(x_{nl}|H_1)} {f_n(x_{nl}|H_0)}
\end{eqnarray}
 and $\overline{R_n^{'}}$ is a single interval. That is only \textit{extremal} or very ``informative" likelihood ratios are transmitted. For the case of dependent observations,
we assume that the censoring scheme in (\ref{eq:cen_general}) is applied with $h_n(\cdot)$ being the likelihood ratio function and $\overline{R_n^{'}}$ being a single-interval\footnote{Finding the optimal censoring region in the case of dependent sensor observations is quite difficult, even with the arguably simplest case of multivariate Gaussian observations \cite{Willett2000}.}. The censoring scheme can be rewritten as
 \begin{eqnarray}
\left \{
  \begin{array}{l l}
    x_{nl} \in R_n, & \quad \text{$u_{nl} = \gamma_n(x_{nl})$ is sent} \\
    x_{nl} \in \overline{R_n}, & \quad \text{nothing is sent}
  \end{array} \right.
\label{eq:cen}
\end{eqnarray}
where 
\begin{eqnarray}
\label{eq:regions}
{R_n} = h_n^{-1}(R_n^{'})  ~\text{and} ~\overline{R_n} = h_n^{-1}(\overline{R_n^{'}})
\end{eqnarray}
If the ratio of the two PDFs is a non-decreasing function in the argument $x_{nl}$, we say that they exhibit the Monotone Likelihood Ratio (MLR) property in $x_{nl}$. For the distributions $f_n(\cdot|H_1)$ and $f_n(\cdot|H_0)$ satisfying the MLR proerty, i.e., $h_n(\cdot)$ is non-decreasing, it can be proved that $\overline{R_n}$ preserves the single interval nature of the censoring region according to its definition in (\ref{eq:regions}).
In this paper, we assume that the observation at each sensor node satisfies the MLR property \footnote{Many families of distributions satisfy the MLR property, such as the one-parameter exponential family.}.
Thus, we have $\overline{R}_n := [t_{n1}, t_{n2}]$, where $t_{n1}$ and $t_{n2}$ are respectively the lower and upper limits of the no-send interval. 

We further define two sets: $\mathbb{C}_l :=\{n: {x_{nl}} \in \overline{R}_n\}$ and $\mathbb{S}_l := \{n: x_{nl} \in R_n\}$ to respectively represent the set of sensors whose observations are censored and the set of sensors whose observations are transmitted at time instant $l$. We use $\mathbf{u}_{\mathbb{S}_l}=\{u_{nl}: n \in \mathbb{S}_l \}$ to denote the set of messages that are transmitted from the sensors to the FC at time instant $l$. 

Assuming ideal sensor-to-FC channels, the FC makes the decision about the true state of nature by combining the messages $\mathbf{u}_{\mathbb{S}}=\{\mathbf{u}_{\mathbb{S}_1}, \dots, \mathbf{u}_{\mathbb{S}_L} \}$ from the sensors with its own observations $\mathbf{x}_0 = \{x_{01}, \dots, x_{0L}\}$. We focus on designing the fusion rule $\gamma_0(\mathbf{u}_{\mathbb{S}},\mathbf{x}_0)$ in the NP framework, assuming that each sensor's censoring scheme is known to the FC. 
The optimal fusion rule in the NP sense maximizes the probability of detection $P_D$ subject to the constraint that $P_F$ is no greater than $\alpha$, i.e., 
\begin{eqnarray}
&&\max_{\gamma_0} P_D(\gamma_0), \nonumber\\
&&\text{subject to} \quad P_F(\gamma_0)\leq \alpha
\end{eqnarray}
where $P_D = P(\gamma_0(\mathbf{u}_{\mathbb{S}},\mathbf{x}_0)=1|H_1)$ and $P_F = P(\gamma_0(\mathbf{u}_{\mathbb{S}},\mathbf{x}_0)=1|H_0)$.  

The design of the fusion rule is considered under two transmission scenarios in which, depending on the mapping from uncensored observations to messages, either continuous or discrete data from sensors is transmitted.

\textbf{Scenario A-C}: Analog censored data is transmitted. Uncensored raw observations are directly transmitted to the FC, i.e., $ \gamma_n(x_{nl}) = x_{nl}$ for $n \in \mathbb{S}_l$. In this case, $\x_{\mathbb{S}_l} = \{x_{nl}: n \in \mathbb{S}_l\}$ is received at time instant $l$, and the fusion of the analog censored messages $\x_{\mathbb{S}} =\{ \x_{\mathbb{S}_1}, \dots, \x_{\mathbb{S}_L} \}$ along with $\x_0$ is considered.

\textbf{Scenario Q-C}: Quantized-censored data is transmitted. Uncensored observations are first quantized by a multilevel finite-range \footnote{In a finite-range quantizer, the input signals that exceed the dynamic range of the quantizer take on the value of the saturation level. The quantizers in this paper all refer to finite range quantizers, for simplicity of presentation we refer to them only as quantizers.} uniform quantizer and then transmitted. Data that fall in the two send-zones $(-\infty,t_{n1})$ and $(t_{n2}, +\infty)$ are respectively quantized by a uniform quantizer with step size $q_n$ that is determined by the number of bits that can be transmitted over the channel. We consider finite-range quantization with negligible saturation error. 
 Any input signal occurring within a given quantization partition is reported at the quantizer output as being at the center of that partition (i.e., the input is rounded-off to the center of the partition). The explicit quantizer output $u_{nl}=\gamma_n(x_{nl})$ is given by:
\begin{equation}
\begin{aligned}
&\gamma_n(x_{nl})= \\
&\left\{
  \begin{array}{l l}
    t_{n1}-L_n q_n+q_n/2, &  x_{nl} \in \left(-\infty,  t_{n1}-(L_n-1) q_n\right) \\
    t_{n1}+q_n\lfloor{\frac{x_{nl}-t_{n1}}{q_n}}\rfloor + q_n/2, &x_{nl} \in\left[t_{n1}-(L_n-1) q_n, t_{n1} \right)\\
    t_{n2}+q_n\lfloor{\frac{x_{nl}-t_{n2}}{q_n}}\rfloor + q_n/2, & x_{nl} \in\left(t_{n2}, t_{n2}+(U_n-1) q_n \right]\\
    t_{n2}+U_n q_n -q_n/2, &x_{nl} \in\left( t_{n2}+(U_n-1) q_n, +\infty \right)
  \end{array} \right.
  \label{eq:quantize}
\end{aligned}
\end{equation}
where $L_n$ and $U_n$ are respectively the number of quantization levels of the two send-zones and $\lfloor x \rfloor$ represents the largest integer that is no greater than $x$. Each quantization partition can be represented by an integer $i_n \in  \{ -L_n, \dots, -1, 0,1, \dots, U_n-1\}$, and corresponds to the value of $k_n(i_n)$ at the quantizer output
\begin{eqnarray}
k_n(i_n)=\left\{
  \begin{array}{l l}
  t_{n1}+i_n q_n+q_n/2, & i_n <0 \\
  t_{n2}+i_n q_n+q_n/2, &i_n \geq 0
  \end{array} \right.
  \label{eq:i_u}
\end{eqnarray}
Thus, the $i_n$-th quantization partition is the set of inputs corresponding to the output value $k_n(i_n)$, i.e., $Q_{i_n}:=\{x_{nl}: \gamma_n(x_{nl}) = k_n(i_n)\}$. In other words, the reception of $u_{nl} = k_n(i_n)$ indicates that the raw observation $x_{nl}$ is in partition $Q_{i_n}$. 


%
%
Fusion rules under these two transmission scenarios will be derived. The design of the fusion rule needs to take into consideration not only the unknown inter-sensor dependence, but also the mechanism of missing data which is the known censoring scheme to achieve better detection performance.

\section{copula-based fusion}
\label{sec:fusion_copula}
In this section, we develop the fusion rules based on copula theory for analog censored observations and quantized-censored observations respectively. Before we proceed, we first briefly introduce the basic concepts of copula theory. 

\subsection{Copula theory}
\label{sec:ct}
Copulas are parametric functions that couple univariate marginal distributions to a valid multivariate distribution. They explicitly model the dependence among random variables, which may have arbitrary marginal distributions. Copula theory is an outcome of the work on probabilistic metric spaces~\cite{Schweizer1983} and a copula was initially defined, on the unit hypercube, as a joint probability distribution for uniform marginals. Their application to statistical inference is possible largely due to Sklar's Theorem, which is stated below without proof~\cite{Nelsen2006}.

\begin{theorem}[Sklar's Theorem] Consider an $N$-dimensional distribution function $F$ with marginal distribution functions $F_1,\ldots,F_N$. Then there exists a copula $C$, such that for all $x_1,\ldots,x_N$ in $[-\infty,\infty]$
\begin{equation}
\label{eq:CopEq1}
F(x_1,x_2,\ldots,x_N) = C(F_1(x_1),F_2(x_2),\ldots,F_N(x_N))
\end{equation}
If $F_n$ is continuous for $1\leq n \leq N$, then $C$ is unique
, otherwise it is determined uniquely on $Ran F_1 \times \ldots \times Ran F_N$ where $Ran F_n$ is the range of cumulative distribution function (CDF) $F_n$. 
Conversely, given a copula $C$ and univariate CDFs $F_1,\ldots,F_N$, $F$ as defined in (\ref{eq:CopEq1}) is a
valid multivariate CDF with marginals $F_1,\ldots,F_N$.
\label{thm:sklar}
\end{theorem}

As a direct consequence of Sklar's Theorem, for continuous distributions, the joint PDF $ f(x_1,\ldots,x_N)$ is obtained by differentiating both sides of (\ref{eq:CopEq1}),
\begin{equation}
\label{eq:CopEq2}
 f(x_1,\ldots,x_N) = \left(\prod_{n=1}^{N}f_n(x_n)\right)c(F_1(x_1),\ldots,F_N(x_N))
\end{equation}
where, $f_n(\cdot)$ is the marginal PDF and $c$ is termed as the copula density given by,
\begin{equation}
\label{eq:CopDens}
 c(\mathbf{v}) = \frac{\partial^NC(v_1,\ldots,v_N)}{\partial v_1,\ldots,\partial v_N}
\end{equation}
where $v_n=F_n(x_n)$. 
As indicated in \cite{Nelsen2006}, there are a finite number of well defined copula families that can characterize most dependence structures. Some of the popular copulas are given in Table
\ref{tab:cop}.
While not explicitly specified in (\ref{eq:CopEq1}) and (\ref{eq:CopEq2}), copula functions contain a \emph{dependence parameter} that quantifies the amount of dependence between the $n$ random variables. We denote the dependence parameter as $\boldsymbol{\phi}$, which, in general, may be a scalar, a vector or a matrix. 

\begin{table*}[t]
\footnotesize
\caption{Some copula functions} \label{tab:cop}
	\centering
		\begin{tabular}{p{5em} | c | c p{18em} | c }
		\hline\hline
 \multicolumn{2}{c|}{ $\:$ } & & &  \\[-1.5ex]
   \multicolumn{2}{c|}{Copulas} & \multicolumn{2}{|c|}{Parametric Form} & Parameter Range\\
		\hline\hline 
& & & &  \\
 & Gaussian &$\Phi_{\Sigma}(\Phi^{-1}(u_1), \ldots, \Phi^{-1}(u_m)),$&$ \Phi_{\Sigma}(\mathbf{x})= \int_0^{\mathbf{x}}\mathcal{N}(\mathbf{x};\boldsymbol{0},\Sigma)d\mathbf{x},\: \mathbf{x} \in \IR^m $ & $\Sigma = [\rho_{ij}], i,j = 1,\ldots,m$\\
\multirow{3}{5em}{Elliptical copulas}& & & $\Phi^{-1}(u) = \underset{x\in\IR}{\inf}\{u\leq\int_0^x\mathcal{N}(x;0,1)dx \}$& $\rho_{ij} \in [-1,1]$\\[2ex]
&Student-$t$ &$t_{\nu,\Sigma}(t_{\nu}^{-1}(u_1), \ldots, t_{\nu}^{-1}(u_m)),$ & $t_{\nu,\Sigma}:$ multivariate Student-$t$ CDF& $\nu:$ degrees of freedom,\\
& & &  $t^{-1}_{\nu}:$ inverse CDF of univariate Student-$t$& $\nu \geq 3$ \\[2ex]
\hline
& & & & \\[-1ex]
  & Clayton &  \multicolumn{2}{c|}{$\left(\sum_{i=1}^m u_i^{-\phi} -1 \right)^{-\frac{1}{\phi}}$}  &  $\phi \in [-1,\infty) \backslash \{0\}$ \\[2ex]
 \multirow{3}{5em}{Archimedean copulas}& Frank   &  \multicolumn{2}{c|}{$-\frac{1}{\phi}\log\left(1 + \frac{\prod_{i=1}^m{[\exp\{-\phi u_i\}-1]}}{\exp\{-\phi\}-1} \right)$}  &
  $ \phi \in \mathbb{R}\backslash \{0\}$ \\[2ex]
 & Gumbel & \multicolumn{2}{c|}{$\exp\left\{- \left(\sum_{i=1}^m (-\ln u_i)^{\phi}\right)^{\frac{1}{\phi}}   \right\}$} & $ \phi \in [1,\infty)$ \\[2ex]
 & Independent & \multicolumn{2}{c|}{$\prod_{i=1}^m u_i$} & -- \\[2ex]
		\hline\hline
		\end{tabular}
\end{table*}

An 
attractive feature of copulas is the invariance property which is given in the following theorem. 
\begin{theorem}[Thm 2.4.3, \cite{Nelsen2006}]
Let X and Y be continuous random variables with copula $C_{XY}$. If $\alpha$ and $\beta$ are strictly increasing on RanX and RanY, respectively, then $C_{\alpha(X)\beta(Y)} = C_{XY}$. Thus, $C_{XY}$ is invariant under strictly increasing transformations of $X$ and $Y$.
\label{inv_pro}
\end{theorem}
This property will be employed when the raw sensor observation $X$ is transformed by a piecewise linear compressor, later in Section \ref{sec:si} of this paper.

An important step in modeling the joint distributions using the copula-based model in (\ref{eq:CopDens}) is how to choose $c(\cdot)$ from a finite set of copulas, say $\mathcal{C}=\{c_m: m=1,\dots,M\}$. As will be shown later, copula model selection is embedded in the GLRT formulation, as well as the estimation of corresponding copula parameter $\boldsymbol{\phi}$.

\subsection{Fusion of analog censored data}

Under \textbf{Scenario A-C}, the joint PDF of received messages $\x_{\mathbb{S}_l}$ and FC's observation $x_{0l}$ under  hypothesis $H_i, (i = 0,1)$ is given as
\begin{eqnarray}
f(\x_{\mathbb{S}_l},x_{0l}|H_i) = {\int \limits_{\prod \limits_{n \in \mathbb{C}_l} \overline{R}_n }f_{\X}(\x_l, x_{0l}|H_i)d \x_{\mathbb{C}_l}}
\label{eq:likelihood_censoring}
\end{eqnarray}
where $\x_l = \{x_{1l}, \dots, x_{Nl} \}$ , $\x_{\mathbb{C}_l}=\{x_{nl}: n\in \mathbb{C}_l\}$ and $f_{\X}(\cdot|H_i)$ denotes the joint density function of all observations $\X:=[X_1,\dots, X_N, X_0]$ from the sensors and the FC under $H_i$. We use $\prod_{n \in \mathbb{C}_l} \overline{R}_n$ to represent the multifold integration regions $\overline{R}_{\mathbb{C}_l\{1\}} \times \dots \times \overline{R}_{\mathbb{C}_l\{|\mathbb{C}_l|\}}$ where $\mathbb{C}_l\{j\}$ is the $j$-th element of $\mathbb{C}_l$ and $|\cdot|$ denotes the cardinality of the set. The expression on the right hand side (RHS) of Equation (\ref{eq:likelihood_censoring}) is the joint PDF of all the observations integrated over the no-send regions of all censoring sensors, which yield the joint PDF of $\{\x_{\mathbb{S}_l},x_{0l}\}$ under $H_i$. The dimension of the integration is $|\mathbb{C}_l|$.

The unknown joint distribution $f_{\X}(\x_l, x_{0l}|H_i)$ can be approximated using a copula density function. According to the copula-based formulation of a joint distribution in (\ref{eq:CopEq2}), the density function in (\ref{eq:likelihood_censoring}) can be approximated by
\begin{eqnarray}
\lefteqn{\hat{f}(\x_{\mathbb{S}_l},x_{0l}|c_i, \boldsymbol{\phi}_i,H_i) } \nonumber \\
&& = {\int \limits_{ \prod \limits_{n \in \mathbb{C}_l} \overline{R}_n }\hat{f}_{\X}(\x_l, x_{0l}|H_i)d \x_{\mathbb{C}_l}} \nonumber \\
&& = \int \limits_{\prod \limits_{n \in \mathbb{C}_l} \overline{R}_n }\prod_{n=0}^N f_n(x_{nl}|H_i) c_i(\x_l,x_{0l}|\boldsymbol{\phi}_i)  d \x_{\mathbb{C}_l}
\label{eq:likelihood_censoring2}
\end{eqnarray} 
with 
\begin{eqnarray}
c_i(\x_l,x_{0l}|\boldsymbol{\phi}_i) = c_i(F_0(x_{0l}|H_i),\dots,F_N(x_{Nl}|H_i)|\boldsymbol{\phi}_i) 
\end{eqnarray}
where $c_i$ denotes the copula density function applied to approximate the dependence structure under hypothesis $H_i$ and $\boldsymbol{\phi}_i$ represents the corresponding dependence parameter. We have used the notation $\hat{f}(\cdot)$ to emphasize that these are approximations. How $c_i$ is selected from a library of copula density function $\mathcal{C}$ and how the parameter $\boldsymbol{\phi}_i$ is estimated according to the data that is available at the FC, will be discussed later in the section.

The Generalized Likelihood Ratio Test (GLRT) at the FC can be obtained based on (\ref{eq:likelihood_censoring2}) 
\begin{eqnarray}
T(\mathbf{x_{\mathbb{S}}},\x_0) = \frac{ \max \limits_{c_1 \in \mathcal{C},\boldsymbol{\phi}_1 } \prod \limits_{l=1}^L {\hat{f}(\x_{\mathbb{S}_l},x_{0l}|c_1, \boldsymbol{\phi}_1,H_1)}}{ \max \limits_{c_0 \in \mathcal{C},\boldsymbol{\phi}_0}\prod \limits_{l=1}^L \hat{f}(\x_{\mathbb{S}_l},x_{0l}|c_0, \boldsymbol{\phi}_0,H_0)} \Gtrless^{H_1}_{H_0} \eta 
\label{eq:GLRT_censored}
\end{eqnarray}
where $\eta$ is the threshold that satisfies the constraint $P_F(\eta) = \alpha$. It should be noted that copula selection and parameter estimation are embedded in the GLRT. 
The best copula $c_i^*$ is  the one that has the highest likelihood score, i.e., 
\begin{eqnarray}
c_i^* = \arg \max \limits_{c_i \in \mathcal{C}}\prod_{l=1}^L {\hat{f}(\x_{\mathbb{S}_l},x_{0l}|c_i, \hat{\boldsymbol{\phi}_i},H_i)}
\end{eqnarray}
where, for any $c_i \in \mathcal{C}$, the corresponding parameter is estimated using maximum likelihood estimation
\begin{eqnarray}
\hat{\boldsymbol{\phi}_i} = \arg \max_{\boldsymbol{\phi}_i} \prod_{l=1}^L {\hat{f}(\x_{\mathbb{S}_l},x_{0l}|c_i, \boldsymbol{\phi}_i,H_i)}
\end{eqnarray}
Since the true dependence model can be very complex and may not be present in the library of candidate copulas $\mathcal{C}$, the best copula $c_i^*$ may still be misspecified. 

When independent observations are assumed across the sensors, the dependence structures under both hypotheses are described by the product copula, i.e., $c_i^* =1, i = 0,1$. In this case, the test statistic in (\ref{eq:GLRT_censored}) reduces to 
\begin{eqnarray}
T(\x_{\mathbb{S}},\x_0) =\prod_{l=1}^L \left[ \prod_{n \in \mathbb{C}_l} \rho_n \prod_{n\in \mathbb{S}_l \cup\{0\} } \frac{f_n(x_{nl}|H_1)}{f_n(x_{nl}|H_0)} \right ]
\label{eq:ind}
\end{eqnarray}
where $\rho_n$ is the likelihood ratio between the two hypotheses when no message is received from sensor $n$, which is given as
\begin{eqnarray}
\rho_n &=&\frac{P(X_{n} \in \overline{R}_n |H_1)}{P(X_{n} \in \overline{R}_n |H_0)} 
\end{eqnarray}
The test statistic in (\ref{eq:ind}) is the same as the one derived under independence assumption in \cite{489500,4456697}. 

Unlike the evaluation of the test statistic in (\ref{eq:ind}), which involves only one-dimensional integrals, the computation of $T(\mathbf{x_{\mathbb{S}}},\x_0)$ in (\ref{eq:GLRT_censored}) for dependent observations involves multiple $|\mathbb{C}_l|$-dimensional integrations due to the existence of spatial dependence. When the probability of censoring becomes higher for each sensor or the number of sensors in the network gets larger, $|\mathbb{C}_l|$ increases, so does the computational complexity of (\ref{eq:GLRT_censored}). 

\subsection{Fusion of quantized-censored data}
In \textbf{Scenario Q-C}, where uncensored observations are quantized before transmission, discrete-valued messages are received at the FC. 
A copula-based rule for fusing these discrete-valued messages and continuous observations of the FC is developed in this subsection. 

Knowing local sensors' censoring schemes, the joint likelihood that the dataset $\mathbf{u}_{\mathbb{S}_l}=\{k_n(i_n): n \in \mathbb{S}_l\}$ is received and $x_{0l}$ is observed at the FC under hypothesis $H_i$ is 
\begin{eqnarray}
\lefteqn{f(\{k_n(i_n):n \in \mathbb{S}_l\}, x_{0l}|H_i) =} \nonumber \\
&&\int \limits_{\prod \limits_{n \in \mathbb{C}_l} \overline{R}_n } \int \limits_{ \prod \limits_{n \in\mathbb{S}_l} Q_{i_n}} f_{\X}(\x_l,x_{0l}|H_i) d\x_l
\label{eq:probability}
\end{eqnarray}
where $d\x_l = d x_{1l} \dots d x_{Nl}$ and recall that $Q_{i_n}$ is the quantization partition corresponds to the output value $k_n(i_n)$.
Eq. (\ref{eq:probability}) is the joint distribution of sensors' and FC's observations integrated over the no-send regions of the censoring sensors and the quantization partitions $Q_{i_n}$ of the transmitting sensors. The unknown joint distribution $f_{\X}(\x_l, x_{0l}|H_i)$ can be approximated using a copula density function. Thus, the probability density function in  (\ref{eq:probability}) can be approximated as follows
\begin{eqnarray}
\lefteqn{\hat{f}(\{k_n(i_n): n \in \mathbb{S}_l\}, x_{0l}|H_i) } \nonumber \\
&&=\int \limits_{\prod \limits_{n \in \mathbb{C}_l} \overline{R}_n } \int \limits_{\prod \limits_{n \in\mathbb{S}_l} Q_{i_n} } \hat{f}_{\X}(\x_l, x_{0l}|H_i) d\x_l \nonumber\\
&&=\int \limits_{\prod \limits_{n \in \mathbb{C}_l} \overline{R}_n } \int \limits_{\prod \limits_{n \in\mathbb{S}_l} Q_{i_n} }\prod_{n=0}^N f_n(x_{nl}|H_i) c_i(\x_l,x_{0l}|\boldsymbol{\phi}_i)d\x_l  \nonumber \\
\label{eq:pmf}
\end{eqnarray}
The dependence of the LHS of (\ref{eq:pmf}) on the copula model $c_i$ and its parameter $\boldsymbol{\phi}_i$ is not specified only for notational simplicity.
Statistical theory of copulas is not applied directly to approximate the joint distribution of the discrete random variables $\{u_1, \dots, u_N\}$ because copulas for discrete marginals are not well defined. Thus, we can only approximate the joint distribution of the continuous random vector $\X$, through which the approximated probability of $\{\mathbf{u}_{\mathbb{S}_l}, x_{0l}\}$ under each hypothesis can be obtained. It has to be noted that an $N$-dimensional integration is involved in (\ref{eq:pmf}).

Based on (\ref{eq:pmf}), the joint distribution of $\mathbf{u}_{\mathbb{S}_l}$ and $x_{0l}$ given $H_i$ can be written as
\begin{eqnarray}
\lefteqn{\hat{f}(\mathbf{u}_{\mathbb{S}_l}, x_{0l}|c_i, \boldsymbol{\phi}_i,H_i)=} \nonumber \\
 & \prod \limits_{\mathbb{S} \in \mathcal{N}}  \prod \limits_{\substack{i_n=-L_n \\ n \in \mathbb{S}_l}}^{U_n-1}
\hat{f}(\{k_n(i_n), n \in \mathbb{S}_l\},x_{0l}|H_i)^{\mathbb{I}_{\mathbb{S}_l=\mathbb{S}}\mathbb{I}_{\mathbf{u}_{\mathbb{S}_l}=\{k_n(i_n):n \in \mathbb{S}_l\}}}\nonumber \\
 \label{eq:ll}
\end{eqnarray}
where $\mathbb{I}_{\{\cdot\}}$ denotes the indicator function. $\mathbb{S}$ represents a subset of $\{1, \dots, N\}$ and $\mathcal{N}$ represents the set consisting of all possible $\mathbb{S}$.

The Generalized Likelihood Ratio Test (GLRT) at the FC can be written as 
\begin{eqnarray}
\label{eq:GLRT_QC}
T(\mathbf{u_{\mathbb{S}}}, \x_0)={\frac{\max \limits_{c_1 \in \mathcal{C},\boldsymbol{\phi}_1 } \prod \limits_{l=1}^L \hat{f}(\mathbf{u}_{\mathbb{S}_l}, \x_0|c_1, \boldsymbol{\phi_1},H_1)}{\max \limits_{c_0 \in \mathcal{C},\boldsymbol{\phi}_0 } \prod \limits_{l=1}^L \hat{f}(\mathbf{u}_{\mathbb{S}_l}, \x_0|c_0, \boldsymbol{\phi}_0,H_0)}}  \Gtrless^{H_1}_{H_0} \eta
\end{eqnarray}
where $\eta$ satisfies the constraint that $P_F(\eta) = \alpha$.

It is noted that evaluation of $T(\mathbf{u_{\mathbb{S}}}, \x_0)$ involves $N$-dimensional integrations, thus the computational complexity increases drastically in the number of sensors. Therefore, we propose computationally efficient approximate fusion rules for both transmission scenarios in the following sections.

\section{Noise-aided fusion of analog censored data}
\label{sec:censored}

An alternative fusion rule for analog censored dependent data is proposed in this section based on substituting unreceived messages with artificial noise. This approach eliminates the necessity of computing multidimensional integrals and is, thus, more computationally efficient at the expense of slight performance loss. 

If the FC receives no signal from sensor $n$ at $l$, then we only know that $x_{nl} \in \overline{R}_n$, since neither the true underlying hypothesis nor the priors of the hypotheses are known. 
Thus, the uninformative prior in (\ref{eq:noise_censored}) is assumed for the distribution of the missing messages. 
\begin{eqnarray}
f(x_{nl}) =
&\left\{ 
\begin{array}{c l}     
    &\frac{1}{t_{n2}-t_{n1}},\quad  x_{nl} \in [t_{n1},t_{n2}]\\
    &0, \quad \text{otherwise}
\end{array}\right.
\label{eq:noise_censored}
\end{eqnarray}
for all $n \in \mathbb{C}_l$. An artificial noise $d_{nl}$ is generated according to the PDF in (\ref{eq:noise_censored}) to represent the unreceived message from sensor $n$ at time instant $l$. 
Let $Z_{n}$ denote the message corresponding to sensor $n$ after the addition of noise, if there is any, whose distribution under hypothesis $H_i$ is given as
\begin{eqnarray}
\lefteqn{f_{Z_n}(z_{n}|H_i)} \nonumber \\
&= \mathbb{I}_{z_{n}\in R_n} f_n(z_{n}|H_i) +\mathbb{I}_{z_{n}\in \overline{R}_n} \frac{P(X_{n}\in \overline{R}_n|H_i)}{t_{n2}-t_{n1}} 
\label{eq:z_pdf}
\end{eqnarray}
for all $n= 1, \dots, N$.  The new set of data consists of the received messages and the generated artificial noise terms, i.e., $\z_l:=\{ \x_{\mathbb{S}_l}, \mathbf{d}_{\mathbb{C}_l}\}$, where $\mathbf{d}_{\mathbb{C}_l} = \{ d_{nl}: n \in \mathbb{C}_l \}$. 

The joint PDF of the data set $\z_l$ and $x_{0l}$ can be approximated using a copula density function 
\begin{eqnarray}
\lefteqn{\hat{f}_{\mathbf{Z},X_0}(\mathbf{z}_l,x_{0l}| c_i,\boldsymbol{\phi}_i, H_i)}\nonumber \\
&&=\prod \limits_{n =1}^N f_{Z_n}(z_{nl}|H_i) f_0(x_{0l}|H_i) c_i(\z_l,x_{0l}|\boldsymbol{\phi}_i) \nonumber \\
&&= \prod_{n \in \mathbb{C}_l}\frac{ P(X_{n} \in \overline{R}_n |H_i)}{t_{n2}-t_{n1}} \prod_{n\in \mathbb{S}_l } f_{Z_n}(z_{nl}|H_i) f_0(x_{0l}|H_i) \nonumber\\
&& \times c_i(\z_l,x_{0l}|\boldsymbol{\phi}_i) 
\label{eq:joint_zx}
\end{eqnarray}
where 
\begin{eqnarray}
\lefteqn{c_i(\z_l,x_{0l}|\boldsymbol{\phi}_i) =}\nonumber\\
& c_i(F_{Z_1}(z_{1l}|H_i),\dots,F_{Z_N}(z_{Nl}|H_i),F_0(x_{0l}|H_i)|\boldsymbol{\phi}_i) 
\end{eqnarray}
Thus, the GLRT can be written as
\begin{eqnarray}
T(\z,\x_0)= {\frac{\max \limits_{c_1 \in \mathcal{C},\boldsymbol{\phi}_1 } \prod \limits_{l=1}^L \hat{f}_{\mathbf{Z},X_0}(\mathbf{z}_l,x_{0l}|c_1, \boldsymbol{\phi}_1,H_1)}{\max \limits_{c_0 \in \mathcal{C},\boldsymbol{\phi}_0} \prod \limits_{l=1}^L \hat{f}_{\mathbf{Z},X_0}(\mathbf{z}_l,x_{0l}|c_0, \boldsymbol{\phi}_0,H_0)}}\Gtrless^{H_1}_{H_0} \eta
\label{eq:interpo_glrt}
\end{eqnarray}
where $\z = \{\z_1, \dots, \z_L\}$. By substituting (\ref{eq:joint_zx}) into the above test, we have a 
\begin{eqnarray}
T(\z,\x_0)&=&\prod_{l=1}^L \left[ \prod_{n \in \mathbb{C}_l} \rho_n \prod_{n\in \mathbb{S}_l } \frac{f_{Z_n}(z_{nl}|H_1)}{f_{Z_n}(z_{nl}|H_0)} \frac{f_0(x_{0l}|H_1)}{f_0(x_{0l}|H_0)}\right] \nonumber\\
&& \times \frac{\max \limits_{c_1 \in \mathcal{C},\boldsymbol{\phi}_1} \prod \limits_{l=1}^L c_1(\z_l,x_{0l}|\boldsymbol{\phi}_1)}{\max \limits_{c_0 \in \mathcal{C},\boldsymbol{\phi}_0} \prod \limits_{l=1}^L c_0(\z_l,x_{0l}|\boldsymbol{\phi}_0)}
\label{eq:interpo_t}
\end{eqnarray}
The first term of the test statistic in (\ref{eq:interpo_t}) which is exactly the same as the test statistic under the independence assumption in (\ref{eq:ind}), corresponds to the differences in the marginal statistics,
while the spatial dependence and interactions are included
in the second term.
 
The test in (\ref{eq:interpo_t}) does not require the computation of multidimensional integrals as the one in (\ref{eq:GLRT_censored}), resulting in great computational efficiency. 
Since the SNR at the FC is decreased due to the addition of artificial noise, detection performance is degraded, but only by a relatively small amount as will be shown later in the simulations.

\section{Noise-aided fusion of quantized-censored data}
\label{sec:quantized-censored}
In this section, a computationally efficient fusion rule for quantized-censored data is proposed based on Widrow's Theorem of quantization \cite{widrow96}. We first briefly introduce Widrow's Theorem of quantization. 

\subsection{A Review of Widrow's Statistical Theorem of Quantization}
According to Widrow\cite{widrow96,widrow56}, quantization of a random variable can be interpreted as the sampling of its PDF.  Also, the PDF of the quantized random variable is the convolution of the original PDF with the PDF of a uniformly distributed random variable, followed by conventional sampling. The PDF of the uniform quantizer output $u_{n}$ can be expressed as:
\begin{eqnarray}
f_{U_n}(x) = \left(f_{W_n}(x) \ast f_{X_n}(x)\right) \sum \limits_{t \in \mathbb{Z}} q_n \delta(x-t q_n - \frac{q_n}{2})
\label{eq:cf_u}
\end{eqnarray}
where $\ast$ represents the convolution operation and $\delta(x)$ is define as
\begin{eqnarray}
\delta(x)=\left\{
\begin{array}{c l}     
    1, & x=0\\
    0, &\text{otherwise}
\end{array}\right.
\label{eq:frnp}
\end{eqnarray}
and $f_{W_n}(x)$ is a uniform PDF as follows
\begin{eqnarray}
f_{W_n}(x)=\left\{
\begin{array}{c l}     
    \frac{1}{q_n}, & -\frac{q_n}{2} \leq x \leq  \frac{q_n}{2}\\
    0, &\text{otherwise}
\end{array}\right.
\label{eq:frnp}
\end{eqnarray}
Uniform quantization introduces two kinds of noise: (a) the additive noise $W_n$ and (b) aliasing error due to sampling. However, if the Characteristic Function (CF) of the input PDF $\varphi_{X_n}(\upsilon)=E[e^{j\upsilon x_n}]$ is band-limited such that $\varphi_{X_n}(\upsilon)=0$ for $|\upsilon|>\frac{\pi}{q_n}$, then in principle the original PDF of the input can be reconstructed from the knowledge of $f_{U_n}(x)$. 

\begin{theorem}[Thm QT1, \cite{widrow96}]
\label{QT1}
If the CF of $X_n$ is ``band-limited", so that 
\begin{equation}
\varphi_{X_n}(\upsilon)=0,  \quad |\upsilon| >\frac{\pi}{q_n}
\end{equation}
then the PDF of $X_n$ can be derived from the PDF of $U_n$.
\end{theorem}
\begin{theorem}[Thm QT2, \cite{widrow96}]
\label{QT2}
If the CF of $X_n$ is ``band-limited", so that 
\begin{equation}
\varphi_{X_n}(\upsilon)=0,  \quad |\upsilon| >\frac{2\pi}{q_n}-\varepsilon
\end{equation}
with $\varepsilon$ positive and arbitrarily small, then the moments of $X_n$ can be calculated from the moments of $U_n$.
\end{theorem}

Noting that the conditions of Theorem \ref{QT1} or Theorem \ref{QT2} are more likely to be satisfied for small quantization step sizes $q_n$, we consider the fusion rule that is suited for high-rate quantization.

\subsection{Computationally efficient fusion of quantized-censored data}
\label{sec:si}
As discussed previously, the high complexity in computing the copula-based GLRT stems from the need for computing multidimensional integrals. We simplify the fusion process by adding controlled noise to the multilevel decisions received at the fusion center based on Widrow's theory.

Given the quantization step size $q_n$ which is determined by the number of bits that can be transmitted over the channel, We first propose a fusion rule that corresponds to a specific censoring region $\overline{R}_n = [0, q_n]$. Then, the fusion rule is generalized to the case of any arbitrary censoring interval. 

For this specific censoring interval, receiving no signal from sensor $n$ implies that the observation of sensor $n$ is in the quantization partition $[0,q_n]$. 
We can reformulate the problem as the one in which each sensor's raw observations are quantized according to the following uniform quantizer and all local quantizer outputs $\{u_{nl}:  n = 1, \dots, N\}$ are transmitted to the FC for decision making.  
\begin{equation}
\begin{aligned}
&\gamma_n(x_{nl})= \\
&\left\{
  \begin{array}{l l}
    -L_n q_n+q_n/2, &  x_{nl} < -(L_n-1) q_n \\
    q_n \lfloor{x_{nl}/q_n}\rfloor+q_n/2, &  -(L_n-1) q_n \leq x_{nl} <U_n q_n\\
    U_n q_n +q_n/2, &  x_{nl} \geq U_n q_n
  \end{array} \right.
  \label{eq:quantize_uni}
\end{aligned}
\end{equation}


According to Widrow's Theorem (\ref{eq:cf_u}), the CF of uniformly quantized data contains repeated and phase-shifted replicas due to the sampling process. We are able to keep the main lobe and filter out the terms due to aliasing in $\varphi_{U_n}(\upsilon)$ using a low pass filter (LPF). To do that, an externally generated noise $d_{nl}$ with PDF $f_{D_n}(\cdot)$  is added to the discrete-valued observation before fusion. Let $d_n$ denote the generated noise and the new observation $z_{nl}$ is
\begin{eqnarray}
z_{nl} = u_{nl} +d_{nl}
\end{eqnarray}
Correspondingly, the CF of the new observation is 
\begin{eqnarray}
\varphi_{Z_n}(\upsilon)=\varphi_{U_n}(\upsilon) \cdot \varphi_{D_n}(\upsilon)
\end{eqnarray}

The CF of noise $D_n$ should be band-limited to play the role of a LPF. A perfect LPF-noise would have a rectangular CF in $-\frac{\pi}{q_n} \leq \upsilon \leq \frac{\pi}{q_n} $ which does not correspond to a valid PDF. Thus, attention needs to be paid while designing $D_n$ such that as little distortion as possible is introduced when transforming the discrete-valued $U_n$ to a continuous variable $Z_n$.

Most distributions that are employed in practice, like the Gaussian, exponential or chi-squared are not perfectly band-limited. This fact does not prevent the application of the quantization theorems if the quantum step size is significantly smaller than the standard deviation.  
If the condition stated in Theorem \ref{QT1} is satisfied, we have the following relationship:
\begin{eqnarray}
Z_n = X_n + W_n + D_n
\end{eqnarray}
Thus, at the FC, the PDF of data $z_{nl}$ under hypothesis $H_i$ can be derived, which is
\begin{eqnarray}
f_{Z_n} (z_{nl}|H_i) = f_{n}(z_{nl}|H_i) \ast f_{W_n}(z_{nl}) \ast f_{D_n}(z_{nl})
\end{eqnarray}

In practice, the quantizer output corresponding to the censoring interval $[0,q_n]$, which is  $q_n/2$, is generated at the FC to represent the missing messages before the addition of LPF-noise. So, $d_{nl}$ is added to $u_{nl}$ for all $n \in \mathbb{S}_l$ and $d_{nl}+q_n/2$ is generated for all $ n\in \mathbb{C}_l$ to obtain the new observation $z_{nl}$.  The joint PDF of the data $\mathbf{z}_l =\{z_{1l}, \dots, z_{Nl}\}$ and $x_{0l}$, which are continuous, can be directly approximated by copula theory as follows 
\begin{eqnarray}
\lefteqn{\hat{f}_{\mathbf{Z},X_0}(\mathbf{z}_l,x_{0l}| c_i, \boldsymbol{\phi}_i, H_i)=}\nonumber \\
&&\prod \limits_{n=1}^N f_{Z_n}(z_{nl}|H_i) f_0(x_{0l}|H_i) c_i(\mathbf{z}_l,x_{0l}|\boldsymbol{\phi}_i)
\end{eqnarray}
Thus, the GLRT can be written as
\begin{eqnarray}
T(\mathbf{\mathbf{z}},\x_0)= {\frac{\max \limits_{c_1 \in \mathcal{C},\boldsymbol{\phi}_1} \prod \limits_{l=1}^L \hat{f}_{\mathbf{Z},X_0}(\mathbf{z_l},x_{0l}|c_1, \boldsymbol{\phi}_1,H_1)}{\max \limits_{c_0 \in \mathcal{C},\boldsymbol{\phi}_0} \prod \limits_{l=1}^L \hat{f}_{\mathbf{Z},X_0}(\mathbf{z_l},x_{0l}|c_0,\boldsymbol{\phi}_0,H_0)}}\Gtrless^{H_1}_{H_0} \eta
\label{eq:lpf_glrt}
\end{eqnarray}
where $\mathbf{z} = \{\mathbf{z_1}, \dots, \mathbf{z_L}\}$. The proposed test in (\ref{eq:lpf_glrt}) is a function of continuous variables only and involve the computation of one-dimensional integrals. Compared with the test statistic in (\ref{eq:ll}) which requires the computation of $N$-dimensional integrals, the noise-aided fusion rule greatly simplifies the test.

The noise-aided fusion rule is designed for the specific censoring interval of $[0,q_n]$ in the above discussion. Next, we generalize the test for the case of an arbitrary no-send region of $[t_{n1},t_{n2}]$. It is worth mentioning that if the conditions for Theorem \ref{QT1} or Theorem \ref{QT2} are satisfied for $X_n$, these conditions are still satisfied even after adding a constant $b$ to $X_n$. This can  be explicitly explained from the expression for the CF of $X_n+ b$:
\begin{eqnarray}
\varphi_{X_n+b}(\upsilon) = e^{j \upsilon b} \varphi_{X_n}(\upsilon)
\end{eqnarray}
Similarly, an arbitrary shift of the quantization transfer characteristic in (\ref{eq:quantize}) will not affect the fulfillment of the conditions of the theorems, either. Thus, without loss of generality, we assume $t_{n1}=0$. 

The problem can be identified as the fusion of non-uniformly quantized dependent data in which all the quantization partitions are of length $q_n$ except for one which is $[0,t_{n2}]$.
 It has been demonstrated that the statistical theory of quantization can also be applied to non-uniformly quantized data \cite{widrow96}.
 We can represent the quantizer by a piecewise linear compressor (we call it a compressor, but whether it is actually a compressor or an expander depends on the ratio $t_{n2}/q_n$), a uniform quantizer and a piecewise linear expander which is the inverse of the compressor. Since it is the probability that data comes from a certain quantization partition that is utilized in our test, we only need to consider the piecewise linear compressor and the uniform quantizer that follows it. 

A piecewise linear compressor is applied to the observations $x_{nl}$ before quantization, whose output $y_{nl}$ is given as 
\begin{eqnarray}
\lefteqn{y_{nl}= g_n(x_{nl})=}\nonumber\\
&\left\{ 
\begin{array}{c l}     
    &x_{nl},\quad  x_{nl} < 0\\
    &\frac{q_n x_{nl}}{t_{n2}}, \quad t_{n1} \leq x_{nl} \leq t_{n2} \\
    &x_{nl}-t_{n2}+q_n, \quad x_{nl}> t_{n2}
\end{array}\right.
\label{eq:compression}
\end{eqnarray}

\begin{remark}
Since $g_n(\cdot)$ is a strictly increasing function, according to the invariance property of copulas in Theorem \ref{inv_pro}, the best copula that approximates the dependence among $[X_{1}, \dots, X_{N},X_0]$, is also the one that describes the dependence structure among $[Y_{1},\dots, Y_{N},X_0]$, with the same dependence parameter. 
\end{remark}

If we can ascertain the band-limitedness of the compressed signal $Y_n$, quantization theory developed for uniform quantization can be applied to the uniform quantizer which follows the compressor. Due to the piecewise linear property of the transformation in (\ref{eq:compression}), the PDF of $Y_n$ contains jumps at the break points of $g_n(\cdot)$. Because of such break points, the CF of $Y_n$ may not be perfectly band-limited.  

\begin{figure}[here]
\centering
\includegraphics[width=\columnwidth,height=!]{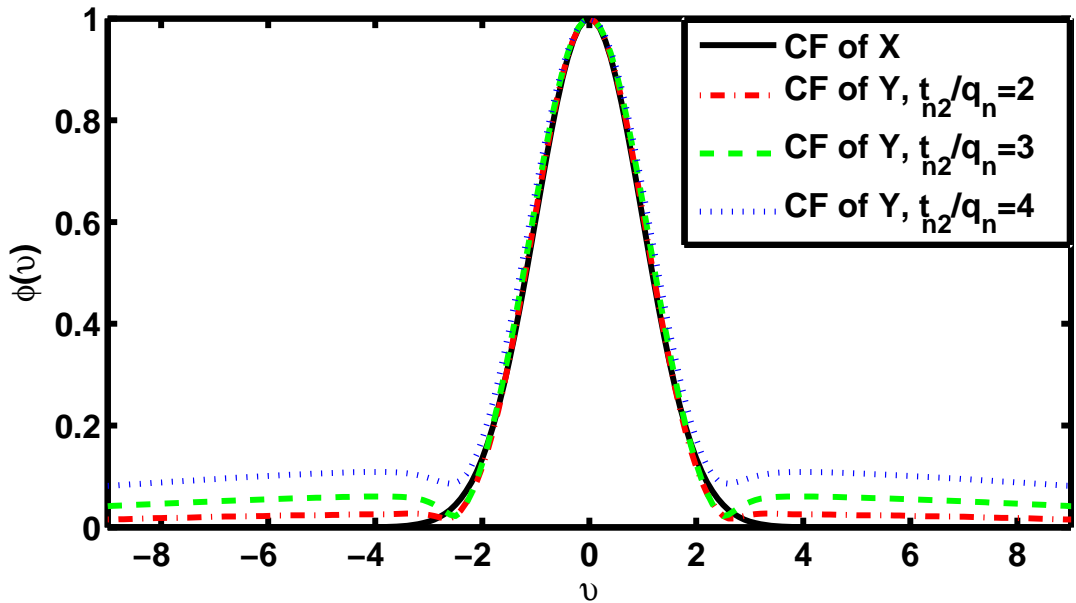}
\caption{Characteristic functions of $X$ and $Y$ }
\label{fig:CF2}
\end{figure}

Figure \ref{fig:CF2} shows that the CF of raw data $X_n$ which is Gaussian distributed and the CF of the compressed signal $Y_n$ with different degrees of compression that is characterized by the ratio between the length of censoring interval and the quantization step size, i.e., $t_{n2}/q_n$. It can be seen that $Y_n$ is not perfectly band-limited. However, when the quantization step size is set to $q_n=1/2$, $\phi(\upsilon) \approx 0$ for $|\upsilon|>\pi/q_n =2\pi$, especially for small degrees of compression ($t_{n2}/q_n<3$). The condition stated in Theorem \ref{QT1} is satisfied very closely, when the length of the censoring interval is comparable to the quantization step size, or in other words, the break point of $g_n(\cdot)$ is not very ``sharp". 

The compressed data, $y_{nl}$, is passed through a uniform quantizer with quantization step size $q_n$ whose output is $u_{nl}=\gamma_n(y_{nl})$. 
This approach successfully transforms the problem of designing computationally efficient fusion rule under arbitrary communication rate constraint to the fusion of uniformly quantized data, to which we already have a noise-aided solution from the previous discussion. 

Before fusion, a LPF-noise $d_{nl}$ is added to each received data $u_{nl}$ to create continuous observation $z_{nl}=u_{nl}+d_{nl}$ and each observation that is not received is replaced by $z_{nl} = q_n/2+ d_{nl}$. 
The new set of observations from sensors to be fused is $\z$ which are continuous.  If the CF of $Y_n$ satisfies the condition of Theorem \ref{QT1} closely and $D_n$ is properly designed as a LPF, we have the following relationship,
\begin{eqnarray}
Z_n = Y_n + W_n + D_n
\end{eqnarray}
At the FC, the PDF of data $z_{nl}$ under hypothesis $H_i$ can be written as 
\begin{eqnarray}
f_{Z_n} (z_{nl}|H_i) = f_{Y_n}(z_{nl}|H_i) \ast f_{W_n}(z_{nl}) \ast f_{D_n}(z_{nl})
\end{eqnarray}
The joint PDF of $\{\mathbf{z},\x_0\}$ can be approximated using a copula density function as in (\ref{eq:joint_zx}) and the test statistic $T(\mathbf{z},\x_0)$ is similar to the one in (\ref{eq:lpf_glrt}) can be derived. 
The fusion rule based on the addition of controlled noise greatly reduces the computational complexity, but at the expense of decreased SNR at the FC. Thus, $D_n$ should be designed to introduce as little distortion as possible while filtering the required signal.

In may practical situations, uniform quantization is not optimal, therefore, it is important to develop models that can accommodate non-uniformly quantized-censored data. It is always possible to represent the nonuniform quantizer by combining a piecewise linear compressor, as the one in (\ref{eq:compression}), a uniform quantizer, and a piecewise linear expander. If the CF of the compressed random variable satisfies the condition for Theorem \ref{QT1}, then the quantization theory for uniformly quantized data can be applied. An example of a floating point quantizer is given in \cite{widrow96}, and for the number of bits used in practice, the conditions required for the quantization theorem are satisfied very closely.

\section{Simulation Results}
\label{sec:sim}

In this section, we provide simulation results to demonstrate the performance of our proposed fusion rules under different settings. We assume that under both hypotheses all sensors' observations are Gaussian distributed as follows
\begin{eqnarray}
&& H_i: X_n \sim \mathcal{N}(\mu_i,\sigma^2), \quad \forall i = 0,1
\end{eqnarray}
for all $n = 1, \dots, N$, where $\mathcal{N}(\mu,\sigma^2)$ denotes the univariate Gaussian distribution with mean $\mu$ and variance $\sigma^2$. It is known that the CF of a Gaussian distribution is a bell shaped function which approximates a band-limited signal. The band-limitedness will be used in the noise-aided fusion under \textbf{Scenario Q-C}. We set $[\mu_0,\mu_1]=[0,0.5]$ and $\sigma=3$ and we assume that under $H_0$, sensors' observations are independently distributed. The FC's observation is distributed according to $\mathcal{N}(0.1,3^2)$. Since the FC is remotely located, we assume that the FC's observation is independent of all sensors' observations under both hypotheses. Identical censoring rate constraints are applied for all sensors, i.e., $\beta_1= \dots =\beta_N = \beta$.
Since we only focus on the fusion aspect of the detection problem for a given local censoring scheme, without loss of generality we assume the censoring region with the lower limit $t_{n1}=0$ and $t_{n2}$ is determined by the following censoring rate constraint 
\[
\int_{0}^{t_{n2}} f_n(x_n|H_0) d x_n = \beta.
\]

We first consider a 2-sensor network, i.e., $N=2$. Under $H_1$, the dependence between the two sensors' observations is generated by a Frank Copula with the corresponding Kendall's $\tau$ being $0.3$. \footnote{Kendall's $\tau$ is a non-parametric rank-based measure of dependence, ranging from $-1$ to $1$. 
Nelsen has proved the following relationship for a copula, $C$, and random variables $X\sim f_X(x), Y\sim f_Y(y)$~\cite[p. 161]{Nelsen2006}: $\tau(\boldsymbol{\phi}) = 4E[C_{\phi}(F_X(x),F_Y(y))] - 1$.}

In \textbf{Scenario A-C}, sensor observations that are not in the censoring region $[0,t_{n2}]$ are transmitted to the FC. The received analog messages are directly used for deciding the true state of nature according to the copula-based GLRT in (\ref{eq:GLRT_censored}), but involving a high computational complexity. In the noise-aided GLRT, the unreceived messages are replaced by randomly generated noise at the FC before fusion. Then the detection is carried out according to (\ref{eq:interpo_t}). In many papers, the inter-sensor dependence is ignored in the fusion of censored data for simplicity. Under independence assumption, the test is conducted according to (\ref{eq:ind}).

In \textbf{Scenario Q-C}, we set the quantization step size $q_n=\sigma/3=1$. Sensor observations that are not in the censoring region $[0,t_{n2}]$ are first quantized and then transmitted to the FC. After receiving the discrete messages, a copula-based GLRT is applied for deciding the true hypothesis according to (\ref{eq:GLRT_QC}). In the noise-aided fusion approach that we proposed for this scenario, our setting of the quantization step size satisfies the conditions in Theorem \ref{QT2} closely.
%
The LPF-noise $D_n$ is designed to be Gaussian distributed with zero mean and variance $\sigma_D = 1$. Distortion is introduced because the CF of Gaussian distribution does not yield an ideal LPF, but is tolerable for our settings. The PDF of  the signal to be fused $Z_n=Y_n+W_n+D_n$, which is nothing but the convolution of the three individual PDFs, is calculated numerically. Having the marginal PDFs, the noise-aided GLRT is conducted to test between the two hypotheses. 

\begin{figure}[here]
\centering
\includegraphics[width=\columnwidth,height=!]{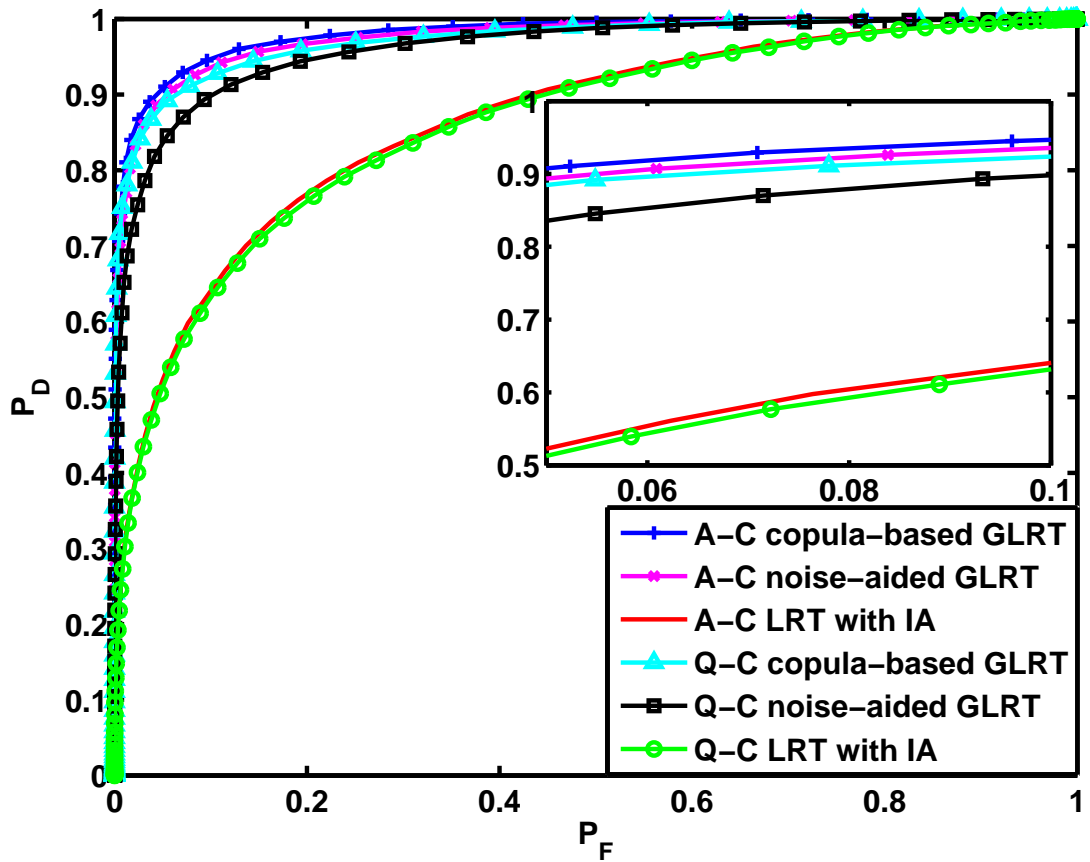}
\caption{ROCs corresponding to different fusion rules in a 2-sensor network with $\beta=0.35$.}
\label{fig:roc}
\end{figure}

We set the copula library, from which the best copula is selected, to be $\mathcal{C}$= \{Gaussian, Gumbel, Frank, Clayton\}, which includes the true generating copula, Frank copula.  The receiver operating characteristic (ROC) curves corresponding to different fusion rules for the given local censoring scheme are depicted in Fig. \ref{fig:roc}.
It can be seen that in both \textbf{Scenario A-C} and \textbf{Scenario Q-C}, our proposed noise-aided GLRTs perform comparably with copula-based GLRTs in which multidimensional integrations are evaluated numerically. Both of our proposed approaches outperform the method under Independence Assumption (IA). The detectors perform better under \textbf{Scenario A-C} compared with those under \textbf{Scenario Q-C}, which is expected since in \textbf{Scenario Q-C} by reducing data transmission through quantization, performance loss is inevitable.

\begin{figure}[here]
\centering
\includegraphics[width=\columnwidth,height=!]{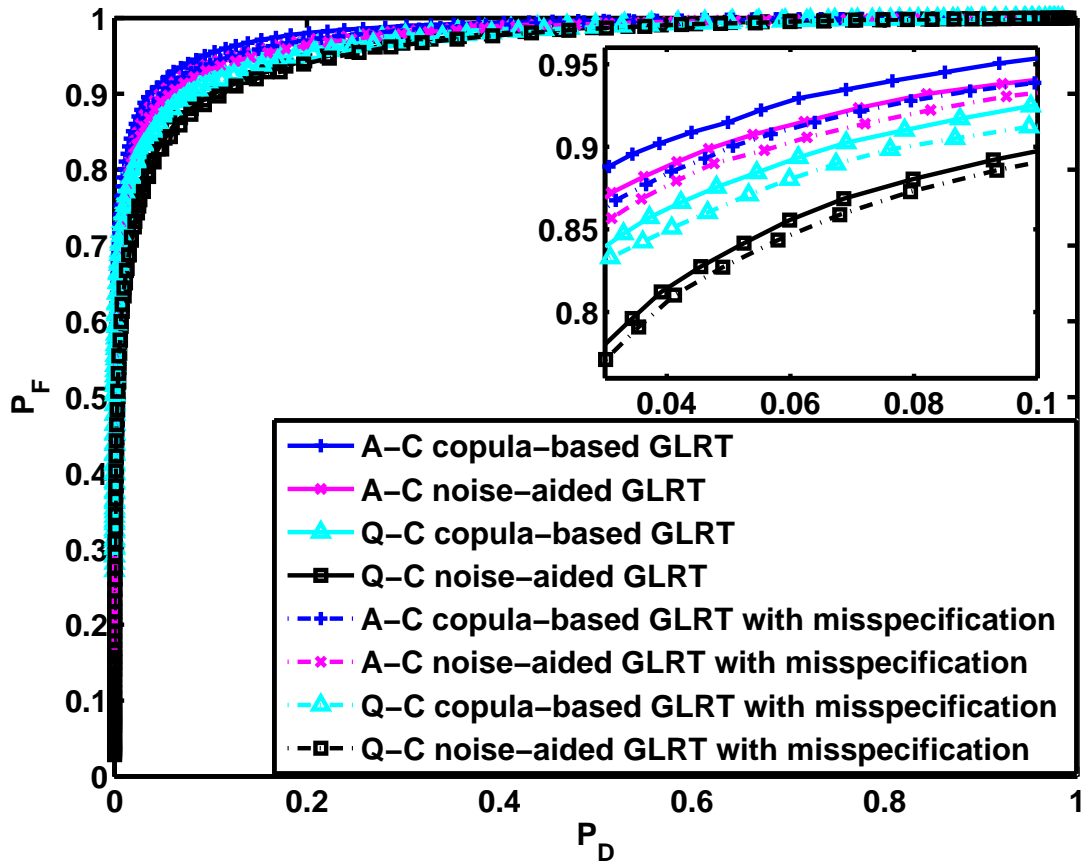}
\caption{ROCs corresponding to different copula libraries with $\beta=0.3$: Frank copula is used to generate the data, GLRT without misspecification corresponds to the case where $\mathcal{C}=$  \{Gaussian, Gumbel, Clayton, Frank\}, while GLRT with misspecification corresponds to the case where $\mathcal{C}=$  \{Gaussian, Gumbel, Clayton\}.}
\label{fig:roc_miss}
\end{figure}

To study the impact of copula misspecification on detection performance, we remove the true copula Frank copula from the copula library. The library of copulas $\mathcal{C}$, including Gaussian copula, Gumbel copula and Clayton copula, does not contain the true copula. Such a setting allows us to examine the performance of copula-based fusion under the unfavorable situation of model misspecification.  
The performance of our proposed fusion rules with copula misspecification is shown in Fig. \ref{fig:roc_miss}. The difference between the two curves corresponding to the detection performance with and without copula misspecification is demonstrated to be negligible in Fig. \ref{fig:roc_miss}. Although the true dependence among sensor observations can be quite complex, a limited number of well defined copula families are able to characterize most dependence structures. Excluding the true copula from the copula library gives us an insight into the detection performance with misspecification in the most unfavorable situation.

\begin{figure}[here]
\centering
\includegraphics[width=\columnwidth,height=!]{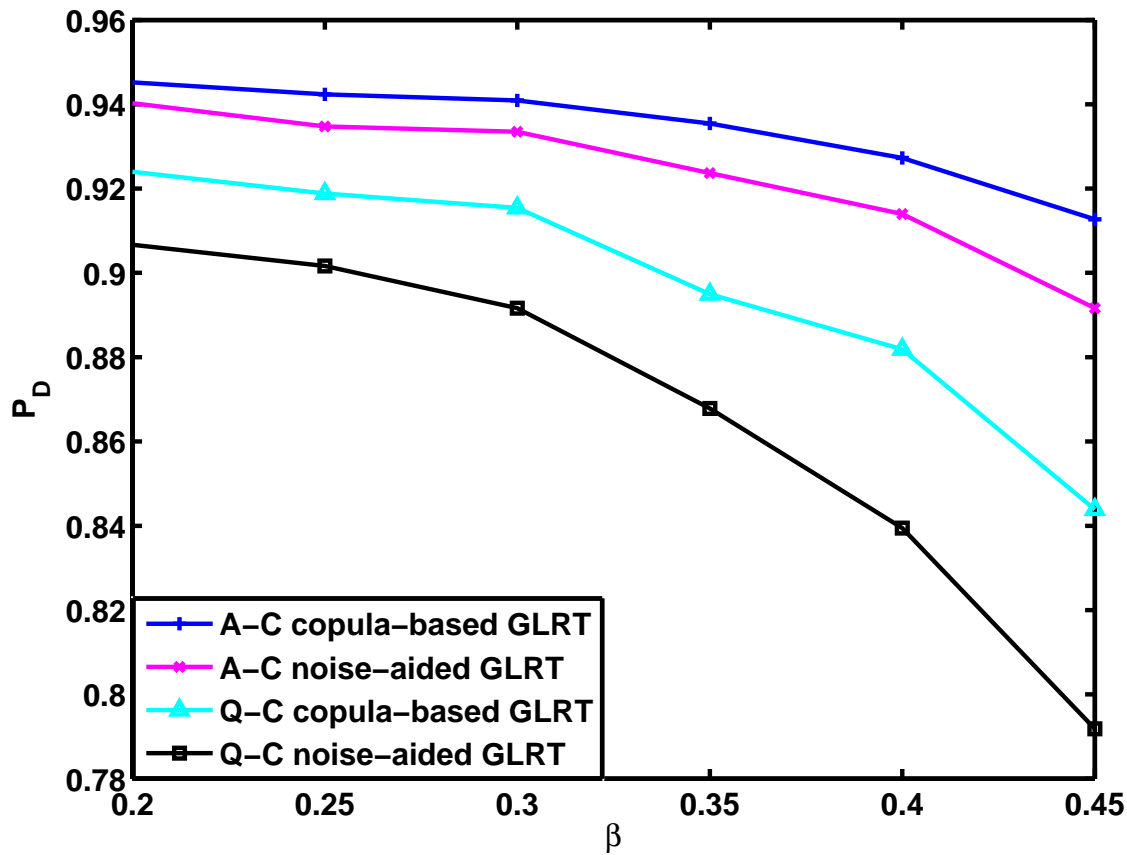}
\caption{$P_D$ as a function of censoring rate $\beta$ under \textbf{Scenario A-C} and \textbf{Scenario Q-C}}
\label{fig:beta_pd}
\end{figure}

The probability of correct detection $P_D$ for a given probability of false alarm $P_F=0.1$ is plotted as a function of censoring rate $\beta$ in Fig. \ref{fig:beta_pd} which captures the tradeoff between detection performance and communication efficiency in a censoring sensor network. It is observed that the performance degrades with increased censoring rate. Under \textbf{Scenario Q-C}, the gap between the performance of the noise-aided GLRT and the copula-based GLRT using brute force integration becomes larger with the increase in censoring rate. A higher censoring rate leads to a higher degree of compression $t_{n2}/q_n$ according to the piecewise-linear compressor defined in (\ref{eq:compression}) in our noise-aided approach. Therefore, the conditions of Widrow's quantization theory are more difficult to satisfy as shown in Fig. \ref{fig:CF2} and the main lobe of the CF of $u_n$ can not be recovered by the process of ``filtering" without noticeable distortion. Thus, attention has to be paid while applying our fusion scheme based on LPF-noise when the censoring rate constraint is high.

\begin{figure}[here]
\centering
\includegraphics[width=\columnwidth,height=!]{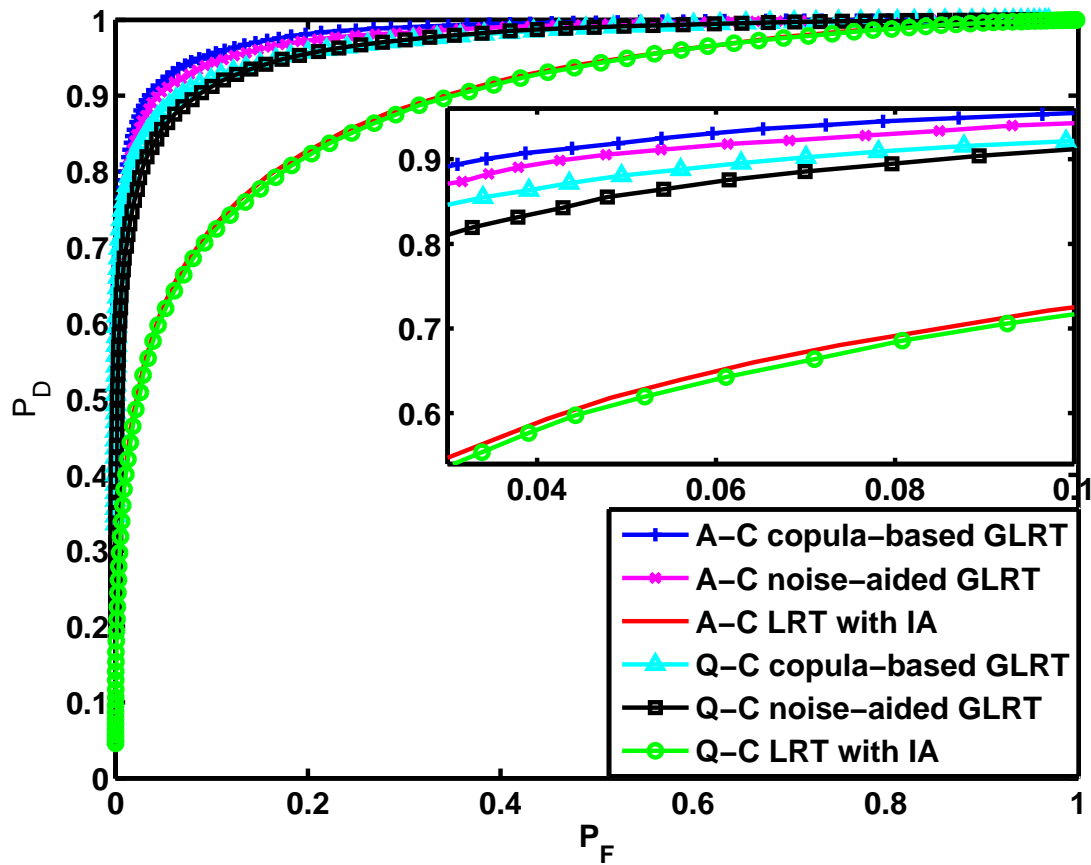}
\caption{ROCs corresponding to different fusion rules in a multi-sensor network with $\beta=0.25$}
\label{fig:ROC_3sensor}
\end{figure}

We also consider a multi-sensor network with $N=3$, where dependence among sensors is generated using a Gaussian copula with the parameter matrix given by 
\[
R = \left[ \begin{array}{ccc}
1 & \rho & \rho \\
\rho & 1 & \rho \\
\rho & \rho & 1 \end{array} \right]
\]
where $\rho=0.25$. In this example, the copula library $\mathcal{C}$ is assumed to include Gaussian copula and t copula. The ROCs corresponding to different fusion approaches in the multi-sensor network are given in Fig. \ref{fig:ROC_3sensor}. As can be seen, the performance of our proposed fusion rules that take the inter-sensor dependence into consideration is better than the test derived under independence assumption (IA). And the noise-aided GLRTs perform comparably with the copula-based GLRTs under both transmission scenarios. With the increase in the number of sensors, computational saving of the noise-aided GLRT which transforms one $N$-dimensional integral to $N$ one-dimensional integrals becomes more significant.

\section{Conclusion}
\label{sec:c}

A binary hypothesis testing problem was considered in a censoring sensor network with spatially dependent observations. Each sensor decides to transmit or to censor based on whether its current observation is ``informative" or not. Two transmission scenarios were considered. In the first one, uncensored observations are transmitted directly to the FC; in the other, uncensored observations are uniformly quantized and then transmitted. Upon the reception of messages from all transmitting sensors, the FC fuses these messages with its own observations to make the final decision. The fusion rules for both analog censored data and quantized-censored data were proposed based on the characterization of unknown spatial dependence using a copula density function. 
The copula-based GLRT for analog censored data involves multidimensional integration, thus is expensive to compute. To address the computational issue, an alternative fusion rule that involves replacing each censored observation with an artificial noise at the FC was proposed. Another computationally efficient fusion rule by injecting controlled noise to the discrete-valued messages was presented to address a similar computational issue with copula-based GLRT for quantized-censored data. 
Simulation results showed that copula-based GLRTs developed here for analog censored data and quantized-censored data and their computationally efficient versions yield significantly superior performance than the ones derived under the independence assumption.
The design of local sensors' censoring strategies and optimal artificial noise to be added at the FC for improved system performance is to be considered in our future work.

\section*{acknowledgment}
This material is based upon work supported by, or in part by, the U. S. Army Research Office under grant number W911-NF-13-2-0040.

\bibliographystyle{IEEEtran}
\bibliography{refs-f12}

\end{document}